\documentclass[10pt,twocolumn,aps,prl,notitlepage,showpacs,floatfix,amsmath,amssymb;twocolumn]{revtex4-1}
\usepackage[latin9]{inputenc}
\usepackage{amsmath}
\usepackage{graphicx}
\usepackage{esint}

\makeatletter

\@ifundefined{textcolor}{}
{%
 \definecolor{BLACK}{gray}{0}
 \definecolor{WHITE}{gray}{1}
 \definecolor{RED}{rgb}{1,0,0}
 \definecolor{GREEN}{rgb}{0,1,0}
 \definecolor{BLUE}{rgb}{0,0,1}
 \definecolor{CYAN}{cmyk}{1,0,0,0}
 \definecolor{MAGENTA}{cmyk}{0,1,0,0}
 \definecolor{YELLOW}{cmyk}{0,0,1,0}
}

\usepackage{float}

\usepackage{epstopdf}
 \usepackage{setspace}

\makeatother

\begin{document}

\title{Probing an effective-range-induced super fermionic Tonks-Girardeau
gas with ultracold atoms in one-dimensional harmonic traps}

\author{Xiao-Long Chen}

\affiliation{Centre for Quantum and Optical Science, Swinburne University of Technology,
Melbourne, Victoria 3122, Australia}

\author{Xia-Ji Liu}

\affiliation{Centre for Quantum and Optical Science, Swinburne University of Technology,
Melbourne, Victoria 3122, Australia}

\author{Hui Hu}

\email{hhu@swin.edu.au}

\affiliation{Centre for Quantum and Optical Science, Swinburne University of Technology,
Melbourne, Victoria 3122, Australia}

\date{\today}
\begin{abstract}
We theoretically investigate an ultracold spin-polarized atomic Fermi
gas with resonant odd-channel ($p$-wave) interactions trapped in
one-dimensional harmonic traps. We solve the Yang-Yang thermodynamic
equations based on the exact Bethe ansatz solution, and predict the
finite-temperature density profile and breathing mode frequency, by
using a local density approximation to take into account the harmonic
trapping potential. The system features an exotic super fermionic
Tonks-Girardeau (super-fTG) phase, due to the large effective range
of the interatomic interactions. We explore the parameter space for
such a fascinating super-fTG phase at finite temperature and provide
smoking-gun signatures of its existence in both breathing mode frequencies
and density profiles. Our results suggest that the super-fTG phase
can be readily probed at temperature at about $0.1T_{F}$, where $T_{F}$
is the Fermi temperature. These results are to be confronted with
future cold-atom experiments with $^{6}$Li and $^{40}$K atoms.
\end{abstract}

\pacs{03.65.Nk, 03.75.Kk, 05.30.-d, 67.85.-d }

\maketitle
The beautiful exactly-solvable models in one dimensional (1D) systems
provide us a better understanding of fascinating low-dimensional quantum
many-body systems in nature \cite{Sutherland2004Book}. Recently,
remarkable experimental progresses in ultracold atoms make it possible
to realize the quasi-1D geometry in laboratory \cite{Richard2003,Moritz2003,Paredes2004,Kinosta004,Moritz2005,Gunter2005,Haller2009,Fang2014},
and therefore pave the way to test a number of exact theoretical predictions
and to confirm the predicted intriguing many-body phenomena \cite{ReviewRMP}.
A well-known example is a 1D Bose gas with strongly repulsive interparticle
interactions, where bosons can not penetrate each other and therefore
their many-body wavefunction resembles that of free spinless fermions
and vanishes whenever two bosons coincide at the same position \cite{Girardeau1960}.
This so-called Tonks-Girardeau (TG) gas has attracted enormous attention
over the past few decades, both experimentally and theoretically \cite{Paredes2004,Kinosta004,Haller2009,ReviewRMP,Kheruntsyan2003,Minguzzi2005,Hu2014,Chen2015}.
To date, evidences of a TG gas have been clearly identified in a number
of experimental observables, including the density profile, momentum
distribution, and collective oscillations \cite{Paredes2004,Kinosta004,Haller2009}.
A highly-excited super-TG Bose gas, which was predicted to occur by
rapidly switching the sign of the interaction strength \cite{Astrakharchik2005,Batchelor2005},
has also been experimentally confirmed \cite{Haller2009}.

In this Letter, we consider the experimental observation of another
fascinating many-body phenomenon, a super fermionic Tonks-Girardeau
(super-fTG) gas. It was predicted to emerge in a spin-polarized Fermi
gas with resonant odd-channel or $p$-wave interactions \cite{Imambekov2010}.
In sharp contrast to the TG or super-TG Bose gas, where the strongly
correlated state is driven by a large scattering length, the super-fTG
gas is caused by a non-negligible effective range of the interparticle
interactions \cite{Imambekov2010}, which is rare in cold-atom experiments.
To explore the realistic parameter space for observing the super-fTG
gas at \emph{finite} temperature, we exactly solve the Yang-Yang thermodynamic
equations for the thermodynamics of the 1D $p$-wave Fermi gas based
on the Bethe ansatz solution \cite{GuanPriviateCommunication}. By
taking into account the external harmonic potential with a trapping
frequency $\omega_{\textrm{ho}}$ via the local density approximation
(LDA) \cite{Dunjko2001}, we calculate the finite-temperature density
distribution of the Fermi cloud. By further using the two-fluid hydrodynamic
theory \cite{Hu2014,Taylor2008,Taylor2009}, we determine the breathing
mode frequency of the low-lying collective oscillations. Clear signatures
of the appearance of a super-fTG gas in these two observables have
been predicted.

\begin{figure}[t]
\centering{} \includegraphics[width=0.48\textwidth]{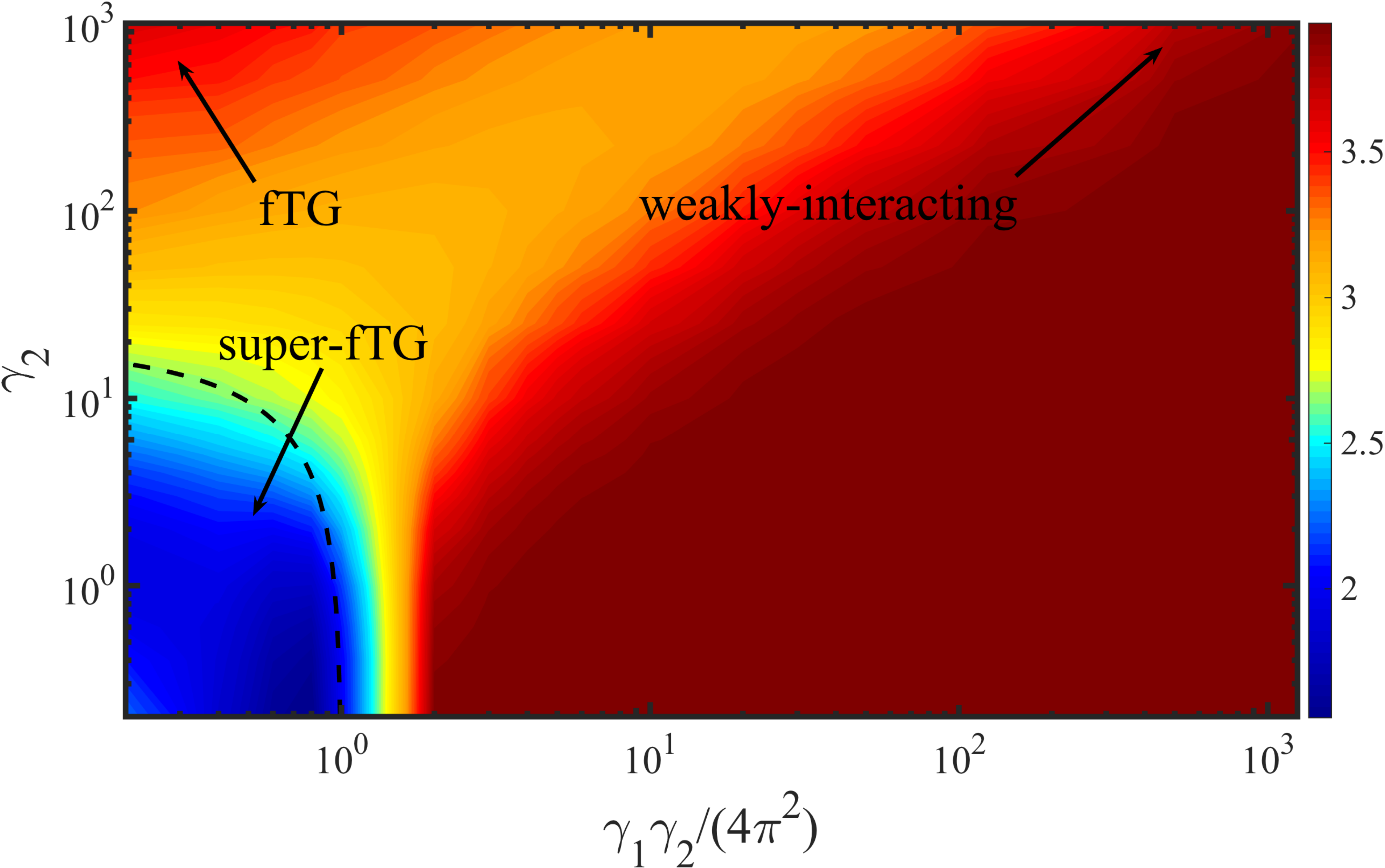}\protect\caption{\label{fig:freq_contour} (color online). Contour plot of the squared
breathing mode frequency $(\omega_{b}/\omega_{\textrm{ho}})^{2}$
as functions of the dimensionless interaction parameters $\gamma_{2}$
and $\gamma_{1}\gamma_{2}/(4\pi^{2})$ (see the text for their definitions)
in the logarithmic scale. The black dashed line is the zero-temperature
analytic result Eq. (\ref{eq:BoundarySuper-fTG}) \cite{Imambekov2010},
indicating the transition into the super-fTG regime, either from the
weakly interacting limit or the strongly interacting fTG limit. We
have taken a typical temperature $T=0.1T{_{\mathrm{F}}}$.}
\end{figure}

Our main result is summarized in Fig. \ref{fig:freq_contour}, where
we report the dependence of the breathing mode frequency on the 1D
scattering length (the horizontal axis) and effective range (the vertical
axis) of the $p$-wave interaction, at a temperature $0.1T_{F}$ that
is typically available in cold-atom experiments. Three distinct regimes
could be clearly identified: a weakly or strongly interacting Fermi
gas with a mode frequency $\omega_{b}\simeq2\omega_{\textrm{ho}}$
and a super-fTG characterized by a much smaller mode frequency at
large effective ranges (i.e., $\gamma_{2}\rightarrow0$). While the
underlying quasiparticles in both weakly and strongly interacting
regimes can be well interpreted in terms of fermions or bosons \cite{ReviewRMP},
the behavior of the compressible super-fTG state is more subtle to
figure out. Therefore, the experimental observation of a super-fTG
gas should provide a new opportunity to understand the challenging
quantum many-body physics.

\textit{1D p-wave atomic Fermi gases}. --- We start by briefly reviewing
the two-particle scattering property in a spin-polarized Fermi gas.
Due to the Pauli exclusion principle, only odd-channel scatterings
are possible, and at the low-energy limit, the $p$-wave scattering
in three-dimensions (3D) is the strongest \cite{LandauLifshitzBookQM,Chin2010}.
Unlike the $s$-wave case, the $p$-wave scattering becomes energy-dependent,
and an effective range of the interaction potential has to be included
in order to regularize the contact interactions. The 3D $p$-wave
scattering is then described by a phase shift $\delta_{p}(k)$: $k^{3}\cot\delta_{p}(k)=-1/w_{1}-\alpha_{1}k^{2}+\mathcal{O}(k^{4}),$
where $k$ is the relative momentum of two colliding atoms, and $w_{1}$
and $\alpha_{1}$ are the scattering volume and effective range, respectively
\cite{LandauLifshitzBookQM,Chin2010}. For $^{6}$Li ($^{40}$K) atoms,
where the $p$-wave resonance occurs near $B_{0}=215.0$ ($198.8$)
G, the effective range $\alpha_{1}$ is about $0.088$ ($0.021$),
in unit of inverse of Bohr radius ($a_{0}^{-1}$) \cite{Zhang2004,Ticknor2004,Schunck2005,Fuchs2008}.
In the quasi-1D geometry considered here, where the transverse motion
is completely suppressed by the strong transverse confinement potential
using a two-dimensional optical lattice \cite{Moritz2003,Kinosta004,Moritz2005,Haller2009},
it is known that the 1D scattering amplitude in the odd channel (denoted
as $p$-wave as well for convenience) takes the form \cite{Granger2004,Pricoupenko2008},
\begin{equation}
f_{p}^{\textrm{odd}}(k)=\frac{-ik}{1/l_{p}+\xi_{p}k^{2}+ik},
\end{equation}
where $l_{p}\approx3a_{\perp}\left[a_{\perp}^{3}/w_{1}-3\sqrt{2}\zeta(-1/2)\right]^{-1}$
and $\xi_{p}=\alpha_{1}a_{\perp}^{2}/3>0$ are the 1D scattering length
and effective range, respectively \cite{Imambekov2010,Granger2004,Pricoupenko2008}.
A confinement induced resonance appears when the 3D scattering length
$w\mathrm{_{1}^{1/3}}$ is comparable to the transverse length $a_{\perp}=\sqrt{\hbar/(m\omega_{\perp})}$,
where $m$ is the atomic mass and $\omega_{\perp}$ is the trapping
frequency of the transverse confinement \cite{Olshanii1998,Granger2004,Pricoupenko2008,Peng2014}.

\textit{Yang-Yang thermodynamic equations.} --- Ignoring the 1D effective
range $\xi_{p}$, a 1D spin-polarized Fermi gas of $N$ atoms is exactly
solvable, owing to the fermion-boson duality \cite{Cheon1998,Cheon1999},
which maps the system into a 1D interacting Bose gas. The latter at
$T=0$ was exactly solved by Lieb and Liniger in 1963 by using the
celebrated Bethe ansatz solution \cite{Lieb1963a,Lieb1963b}. The
finite-temperature thermodynamics of a 1D Bose gas was also solved
a few years later by Yang and Yang, using an approach that is now
commonly referred to as the Yang-Yang thermodynamic equations \cite{Yang1969}.
In the presence of a non-negligible effective range $\xi_{p}\neq0$,
a similar Bethe ansatz for all the many-body wave-functions $\Psi$
can be constructed, by imposing a Bethe-Peierls boundary condition,
$\lim_{x\rightarrow0^{+}}(1/l_{p}+\partial_{x}-\xi_{p}\partial_{x}^{2})\Psi(x=\left|x_{i}-x_{j}\right|;X)=0$
whenever two particles at $x_{i}$ and $x_{j}$ approach each other
\cite{Imambekov2010}, which leads to a set of coupled equations,
\begin{eqnarray}
e^{ikL} & = & \prod_{q}\frac{\xi_{p}\left(k-q\right)^{2}-1/|l_{p}|+i\left(k-q\right)}{\xi_{p}\left(k-q\right)^{2}-1/|l_{p}|-i\left(k-q\right)}.\label{eq:BetheAnsatz}
\end{eqnarray}
Here, the quasi-momenta $k$ and $q$ take $N$ discrete values, and
$L$ is the length of the system under a periodic boundary condition.
We consider only the attractive case $l_{p}<0$, since otherwise the
energy does not have a proper thermodynamic limit \cite{Imambekov2010}.
At $T=0$, the ground state of the system has been solved by Imambekov
\textit{et al.}, by seeking the lowest energy state of Eq. (\ref{eq:BetheAnsatz})
\cite{Imambekov2010}.

At finite temperature, the Yang-Yang thermodynamic equations of a
polarized Fermi gas with a finite $\xi_{p}$ can also be similarly
derived \cite{GuanPriviateCommunication}. In the thermodynamic limit
($N\rightarrow\infty$ and $L\rightarrow\infty$), they take the exactly
same form as that of bosons \cite{Imambekov2010,GuanPriviateCommunication},
except a new kernel function, 
\begin{equation}
\mathcal{K}(k,q)=\frac{2|l_{p}|\left[1+|l_{p}|\xi_{p}\left(k-q\right)^{2}\right]}{\left[1-|l_{p}|\xi_{p}\left(k-q\right)^{2}\right]^{2}+l_{p}^{2}\left(k-q\right)^{2}}.
\end{equation}
To be more explicit, the Yang-Yang thermodynamic equations are given
by ($k_{B}=1$) \cite{Yang1969},
\begin{eqnarray}
 &  & \epsilon\left(k\right)=\frac{\hbar^{2}k^{2}}{2m}-\mu-\frac{T}{2\pi}\int_{-\infty}^{\infty}\mathcal{K}\left(k,q\right)\ln\left[1+e^{-\frac{\epsilon\left(q\right)}{T}}\right]dq\nonumber \\
 &  & 2\pi\rho\left(k\right)\left[1+e^{\frac{\epsilon\left(k\right)}{T}}\right]=1+\int_{-\infty}^{\infty}\mathcal{K}\left(k,q\right)\rho\left(q\right)dq,
\end{eqnarray}
where $\epsilon\left(k\right)$ may be interpreted as the quasi-particle
excitation energy relative to the chemical potential $\mu$, and $\rho(k)$
is the quasi-momentum distribution function normalized according to
$n=N/L=\int\rho(k)dk$. Once the Yang-Yang equations are solved, all
the thermodynamic variables, for example, the total energy and pressure
of the system can be calculated straightforwardly, by using $E=[\hbar^{2}L/(2m)]\int k^{2}\rho(k)dk$
and $P=[T/(2\pi)]\int\ln[1+\exp(-\epsilon(k)/T)]dk$, respectively
\cite{Yang1969}.

To take into account the slowly-varying harmonic trapping potential
in the longitudinal $x$-direction $V_{T}(x)=m\omega_{\textrm{ho}}^{2}x^{2}/2$,
which is necessary to keep atoms from escaping \cite{Richard2003,Moritz2003},
we apply the LDA approximation \cite{Dunjko2001}. This amounts to
setting $\mu[n(x)]=\mu_{0}-V_{T}(x)$, where $n(x)$ is the local
density that is to be inversely solved once we know the relation $\mu(n)$
from the Yang-Yang equations and $\mu_{0}$ is a global chemical potential
to be determined by using $\int n(x)dx=N$ \cite{Dunjko2001}. In
our numerical calculations, two dimensionless interaction parameters
related to the 1D scattering length $|l_{p}|$ and effective range
$\xi_{p}$ are needed. Therefore, we define respectively $\gamma_{1}\equiv1/(n_{\textrm{F}}|l_{p}|)$
and $\gamma_{2}\equiv1/(n_{\textrm{F}}\xi_{p})$, using the peak density
of a zero-temperature ideal Fermi gas at the same trap, $n_{\textrm{F}}=\sqrt{2N}/(\pi a_{\textrm{ho}})$,
where $a_{\textrm{ho}}=\sqrt{\hbar/(m\omega_{\textrm{ho}})}$ is the
characteristic length along the $x$-axis. To be specific, we consider
a polarized Fermi gas of $N=100$ $^{6}$Li atoms under the quasi-1D
confinement with $\omega_{\perp}=2\pi\times200$ kHz and $\omega_{\textrm{ho}}=2\pi\times200$
Hz, leading to a 1D effective range $\xi_{p}=1.58a_{\textrm{ho}}$
and $\gamma_{2}\simeq0.14$.

\textit{Two-fluid hydrodynamics.} --- We are particularly interested
in the low-lying collective oscillations of the Fermi cloud, which
are well described by a two-fluid hydrodynamic theory. At finite temperature,
it takes the following form \cite{Hu2014,Taylor2008,Taylor2009},
\begin{equation}
m\left(\omega^{2}-\omega_{\textrm{ho}}^{2}\right)nu\left(x\right)+\frac{\partial}{\partial x}\left[n\left(\frac{\partial P}{\partial n}\right)_{\bar{s}}\frac{\partial u\left(x\right)}{\partial x}\right]=0,\label{eq:HydrodynamicEquation}
\end{equation}
where $u(x)$ is a displacement field characterizing the oscillation
at frequency $\omega$, and the derivative of the local pressure $P$
with respect to the density $n$ should be taken at the constant local
entropy per particle $\bar{s}=s/n$. In free space, the displacement
field $u(x)$ takes a plane-wave solution with the dispersion $\omega=cq$,
with a sound velocity $c=\sqrt{(\partial P/\partial n)_{\bar{s}}/m}$.
In the presence of the confining harmonic traps, the low-lying collective
modes can be solved by using a polynomial ansatz and to a good approximation,
the breathing mode frequency $\omega_{b}$ is given by \cite{Hu2014},
\begin{equation}
\omega_{b}^{2}=\omega_{\textrm{ho}}^{2}+\frac{\int_{-\infty}^{\infty}\left[\left(\partial P/\partial n\right)_{\bar{s}}/m\right]n\left(x\right)dx}{\int_{-\infty}^{\infty}x^{2}n\left(x\right)dx},\label{eq:GeneralizedSumRules}
\end{equation}
which can be regarded as a finite-temperature generalization of the
well-known sum-rule approach \cite{Dalfovo1999,Pitaevskii2003Book}.

\begin{figure}
\centering{}\includegraphics[width=0.48\textwidth]{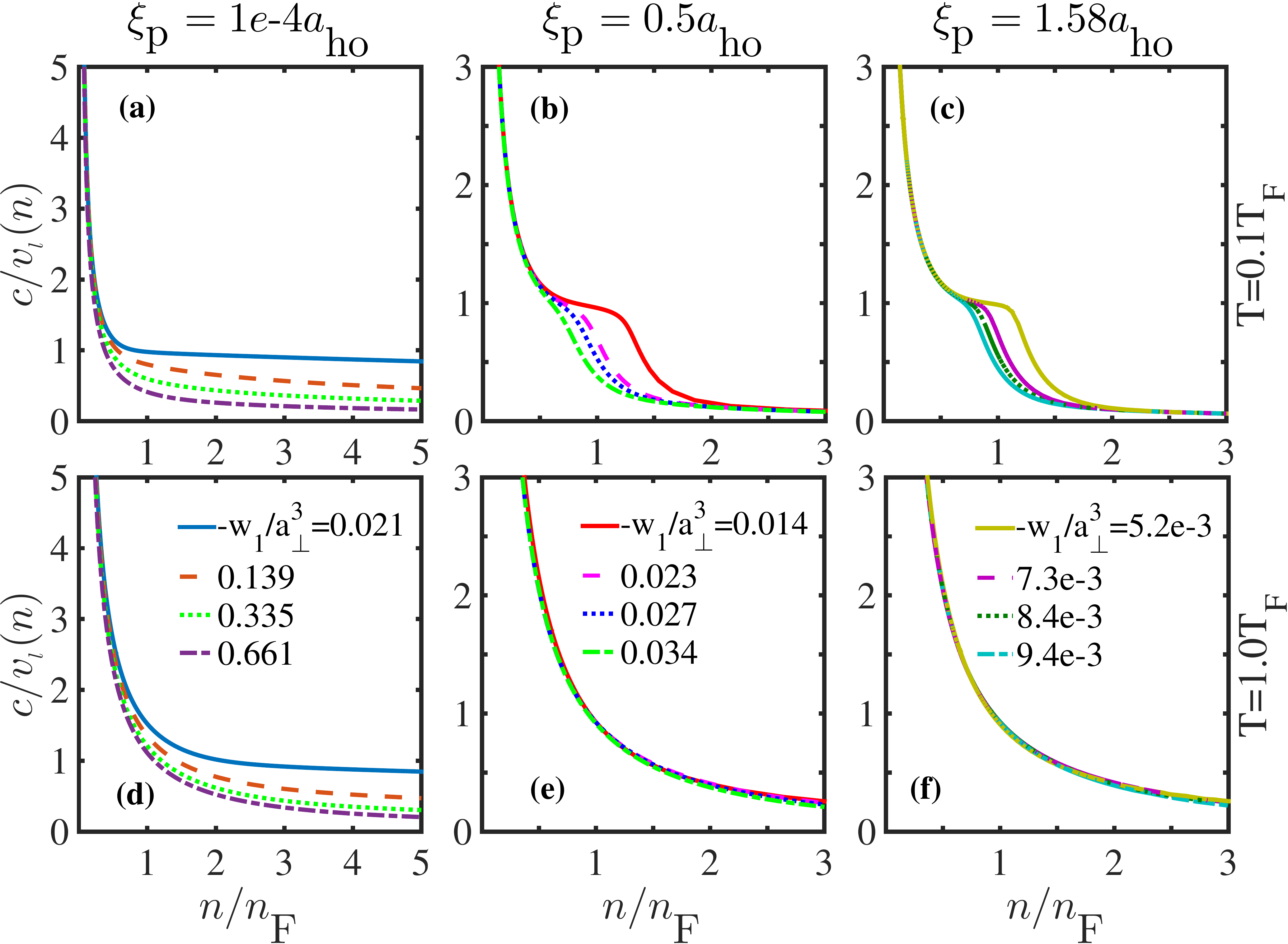}\protect\caption{\label{fig:soundV} (color online). Local sound velocity $c/v_{_{l}}(n)$
as a function of the local density $n/n_{_{\mathrm{F}}}$ for three
sets of effective ranges $\xi_{p}=10^{-4}a_{\textrm{ho}}$ {[}(a),
(d){]}, $0.5a_{\textrm{ho}}$ {[}(b), (e){]}, and $1.58a_{\textrm{ho}}$
{[}(c), (f){]} at low temperature $T=0.1T{_{\mathrm{F}}}$ (upper
panel) and high temperature $T=T{_{\mathrm{F}}}$ (lower panel). In
each subplot, the four curves at different interaction parameters
$-w_{1}/a_{\perp}^{3}$ correspond to the four highlighted points
in the curves of the squared breathing mode frequency, as shown in
the insets of Fig. \ref{fig:density}. Here, $v_{_{l}}(n)=\pi\hbar n/m$
is the local Fermi velocity and $T_{\textrm{F}}=N\hbar\omega_{\textrm{ho}}$
is the Fermi temperature.}
\end{figure}

\textit{Sound speed in free space.} --- In Fig. \ref{fig:soundV},
we present the density dependence of the local sound velocity at some
selected interaction parameters and at a typical experimental temperature
(upper panel, $T=0.1T_{F}$) as well as at a high temperature (lower
panel, $T=T_{F}$). We note that, while the 1D effective range is
directly measured in units of the harmonic oscillator length $a_{\textrm{ho}}$,
the 1D scattering length is indirectly characterized using the 3D
scattering volume $-w_{1}/a_{\perp}^{3}$ for the convenience to make
contact with experiments, where the magnetic field dependence of $\textrm{\ensuremath{\textrm{\ensuremath{w_{1}(B)}}}}$
is known \cite{Zhang2004,Ticknor2004,Schunck2005,Fuchs2008}.

In the case of a negligible effective range (i.e., Figs. \ref{fig:soundV}(a)
and \ref{fig:soundV}(d)), the sound velocity in units of the Fermi
velocity, $c/v_{l}(n)$, decreases monotonically with increasing local
density $n$. The sharp decrease at low density can be understood
from the fermion-boson duality \cite{Cheon1999}. At sufficient small
density $n\ll n_{_{\mathrm{F}}}$, the Fermi cloud lies in the fTG
regime of 1D strongly interacting fermions and is equivalent to a
weakly interacting Bose gas \cite{Cheon1999}, in which the sound
velocity $c\sim n^{1/2}$ \cite{Dalfovo1999,Pitaevskii2003Book}.
As a result, we find that $c/v_{_{l}}(n)\sim(|l_{p}|n)^{-1/2}$, which
quantitatively accounts for the observed rapid decrease. Instead,
at large density ($n\gg n_{_{\mathrm{F}}}$), the sound velocity saturates
to a value that strongly depends on the scattering volume.

The density dependence of the sound velocity changes qualitatively,
when the effective range comes into play (see Figs. \ref{fig:soundV}(b)
and \ref{fig:soundV}(c)). At low temperature, in addition to the
rapid decrease at low density, a plateau develops at the moderate
density $n\sim n_{_{\mathrm{F}}}$, whose structure sensitively relies
on the scattering volume $w_{1}$. By further increasing density,
there is another rapid decrease. The sound velocity finally approach
an asymptotically value that seems less sensitive to the scattering
volume. The observed plateau in $c/v_{_{l}}(n)$ at non-negligible
effective ranges might be interpreted as the emergence of the exotic
super-fTG phase. Although the plateau is washed out at sufficiently
large temperature, as shown in Fig. \ref{fig:soundV}(e) and \ref{fig:soundV}(f),
it could be measured experimentally by creating a density dip at the
trap center and then observing its propagation, following the routine
established for a unitary Fermi gas \cite{Joseph2007}.

\begin{figure}[t]
\centering{}\includegraphics[width=0.48\textwidth]{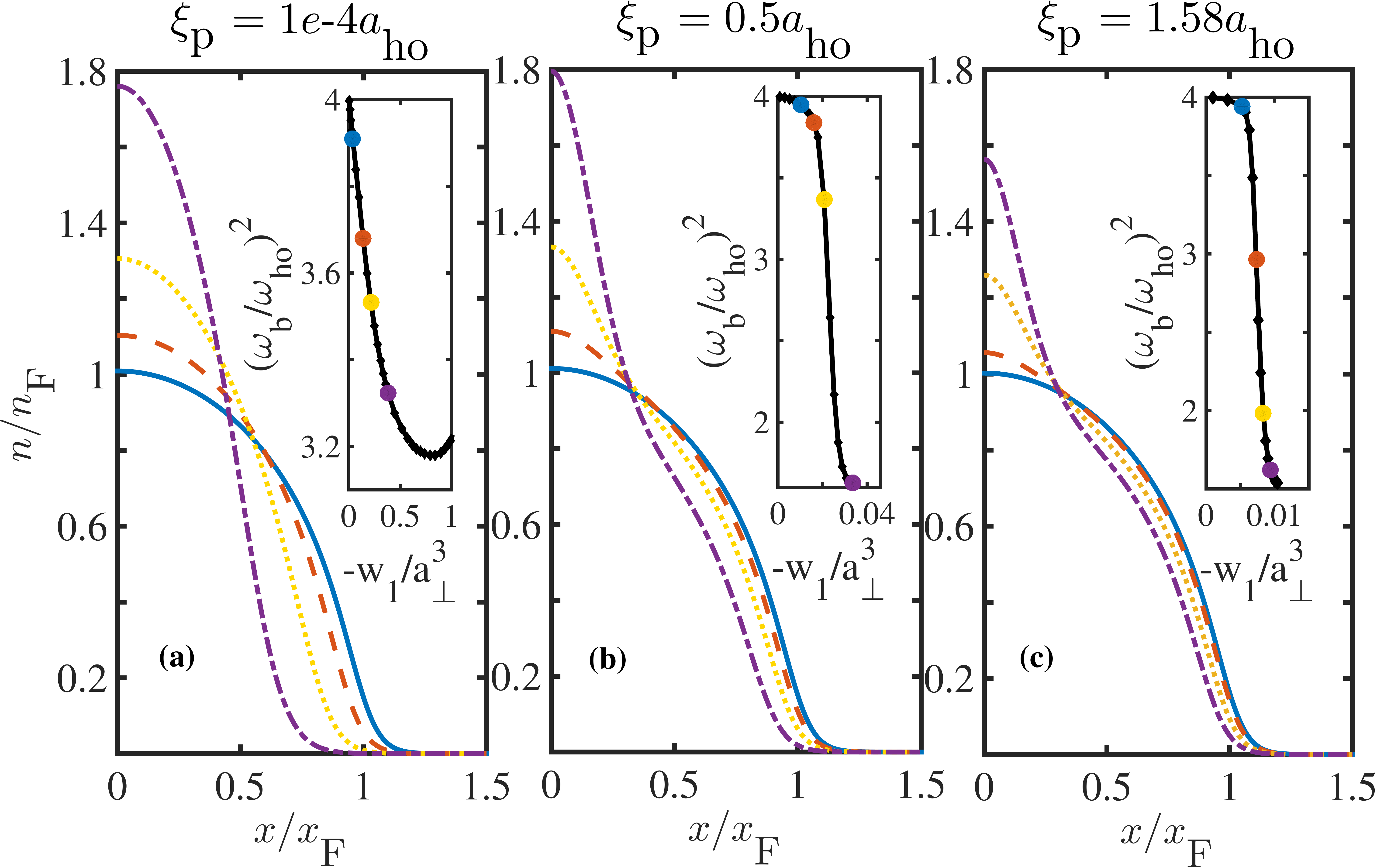}\protect\caption{\label{fig:density} (color online). Density profiles at different
effective ranges: (a) $\xi_{p}=10^{-4}a_{\textrm{ho}}$, (b) $0.5a_{\textrm{ho}}$,
and (c) $1.58a_{\textrm{ho}}$, at $T=0.1T{_{\mathrm{F}}}$. The inset
shows the squared breathing mode frequency as a function of the interaction
parameter $-w_{1}/a_{\perp}^{3}$. In each subplot, the interaction
parameters of colored curves can be read from the highlighted points
with the same color. They are also explicitly indicated in Fig \ref{fig:soundV}.
Here, $x_{\textrm{F}}=\sqrt{2N}a_{\textrm{ho}}$ is the radius of
an ideal trapped Fermi gas at zero temperature.}
\end{figure}

\textit{Density profile at $T\neq0$.} --- Fig. \ref{fig:density}
reports the finite-temperature density distributions at different
effective ranges, and at certain values of the interaction parameter
$-w_{1}/a_{\perp}^{3}$ as illustrated by differently colored curves.
The results at a negligible effective range in Fig. \ref{fig:density}(a)
may again be understood from the fermion-boson duality \cite{Cheon1999}.
At a weak interaction parameter (i.e., the blue line), the profile
is simply an ideal Fermi gas distribution. When the interaction becomes
more attractive, as described by the Cheon-Shigehara (CS) model \cite{Cheon1999},
the Fermi cloud is dual to an interacting Bose gas with an appropriate
repulsion strength $\propto\left|l_{p}\right|^{-1}\propto\left|w_{1}\right|^{-1}$.
Thus, the profile becomes narrower and the peak density is higher,
behaving exactly the same as a 1D Bose gas \cite{Dunjko2001}.

In the presence of sizable effective ranges, as shown in Figs. \ref{fig:density}(b)
and \ref{fig:density}(c) for $\xi_{p}=0.5a_{\textrm{ho}}$ ($\gamma_{2}\simeq0.44$)
and $\xi_{p}=1.58a_{\textrm{ho}}$ ($\gamma_{2}\simeq0.14$), the
shape of the density profile is greatly altered, even by a small increase
in the interaction parameter $-w_{1}/a_{\perp}^{3}$. The peak density
increases significantly, probably due to the enhanced attraction by
the finite effective-range. Furthermore, the profile at large effective
range clearly shows a bimodal distribution. The dramatic change in
the density distribution comes along with a sharp decrease in the
breathing mode frequency, as reported in the two insets, which we
shall now discuss in greater detail. We note that, at zero temperature,
similar changes have been observed by Imambekov and co-workers \cite{Imambekov2010}.

\begin{figure}[t]
\centering{}\includegraphics[width=0.48\textwidth]{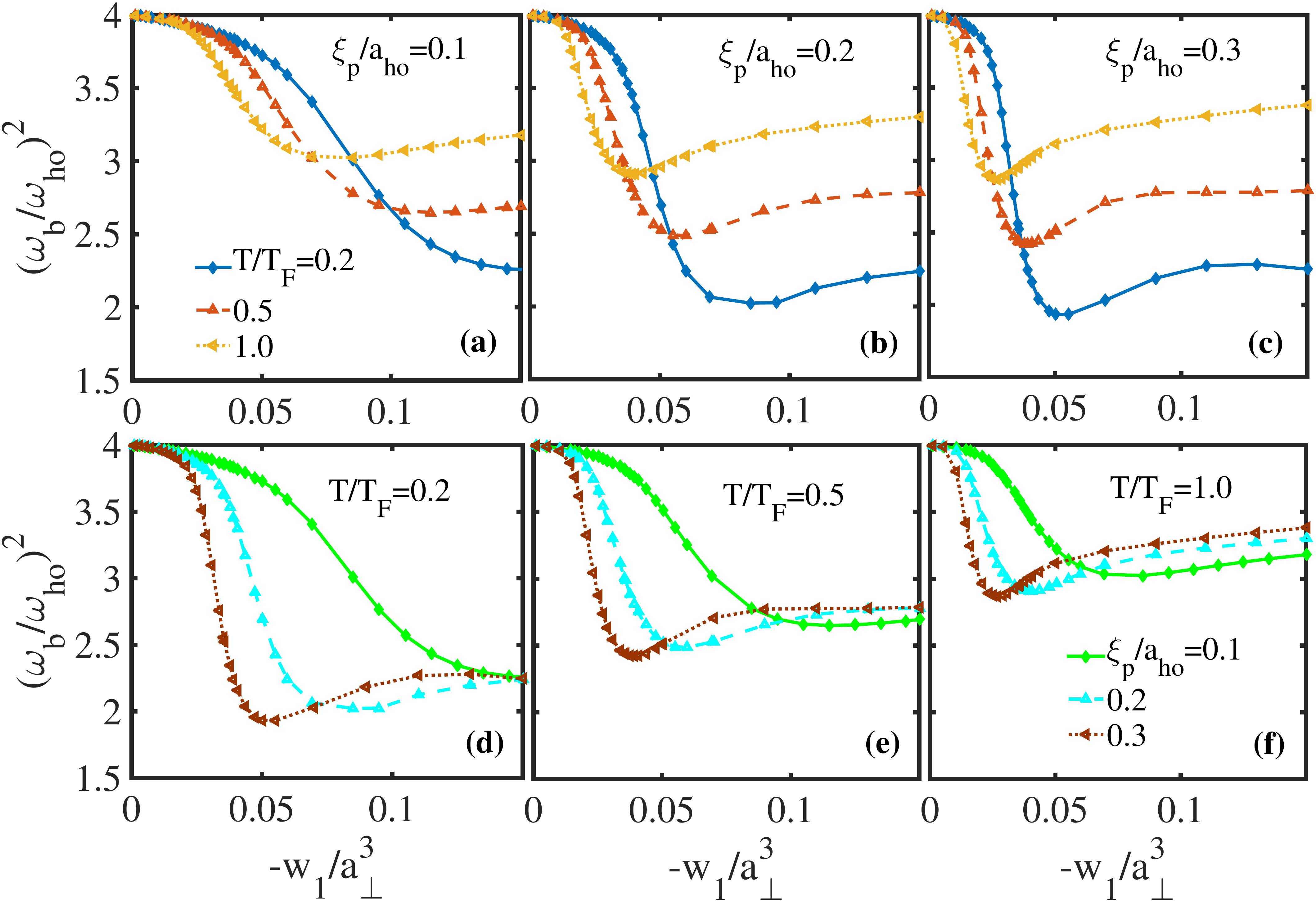}\protect\caption{\label{fig:freq_T_xi} (color online). The squared breathing mode
frequencies $(\omega_{b}/\omega_{\textrm{ho}})^{2}$ as a function
of the interaction parameter $-w_{1}/a_{\perp}^{3}$, at different
effective ranges (upper panel (a)-(c)) or at different temperatures
(lower panel, (d)-(e)), as indicated.}
\end{figure}

\textit{Breathing mode at $T\neq0$.} --- With the sound velocity
$c(x)$ and density distribution $n(x)$ at hand, we calculate straightforwardly
the finite-temperature breathing mode frequency using the generalized
sum-rule approximation Eq. (\ref{eq:GeneralizedSumRules}). The measurement
of collective excitations proved to be a powerful and convenient way
to characterize possible new quantum states of matter arising from
the intriguing effects of interatomic interactions \cite{Dalfovo1999,Hu2004}.
In an interacting 1D Bose gas, the transition from the weakly-interacting
regime to the TG regime for impenetrable bosons is characterized by
a nontrivial but smooth evolution of the squared breathing mode frequency,
which starts at about $4$ in the ideal gas limit, decreases to $3$
at the mean-field Gross-Pitaevskii regime, and then increases back
to $4$ in the TG limit \cite{Haller2009,Chen2015}. At a sufficiently
small effective range, we have checked the above smooth evolution
of the mode frequency, as anticipated from the fermion-boson mapping
\cite{Cheon1999}.

Fig. \ref{fig:freq_T_xi} presents the breathing mode frequency at
some finite effective ranges and finite temperatures. Typically, we
find that the mode frequency experiences a sudden drop at a certain
critical value of $-w_{1}/a_{\perp}^{3}$, after which the frequency
slowly increases. This sudden change could be viewed as a clear signature
of the appearance of the super-fTG phase. It is readily seen that
this sudden-drop feature is enhanced by a large effective range, which
also leads to a minimum squared frequency as small as $2\omega_{\textrm{ho}}^{2}$.
A finite temperature tends to significantly lift the minimum frequency.
However, the sudden-drop structure is merely unchanged.

A typical phase diagram of the 1D polarized Fermi cloud can then be
summarized, as shown earlier in Fig. \ref{fig:freq_contour} for an
experimentally reachable temperature $T=0.1T_{F}$. At large enough
$\gamma_{2}$ (i.e. negligible effective range), the squared frequency
$(\omega_{b}/\omega_{\textrm{ho}})^{2}$ of the system shows a reentrant
behavior, valuing about $4$ in the weakly-interacting limit, changing
to $3$ at the intermediate regime and finally returning back to $4$
in the fTG limit, as a result of the duality to a 1D Bose gas. In
contrast, at sufficiently small $\gamma_{2}$ (i.e. sizable effective
range), with increasing $|l_{p}|$ or decreasing $\gamma_{1}$, the
frequency ratio loses the reentrant behavior and drops sharply to
a much lower value at a critical interaction parameter, thereby signifying
the phase transition to the super-fTG phase. At low temperature, the
critical interaction parameter may be estimated from the ground-state
energy \cite{Imambekov2010}, 
\begin{equation}
\left[\frac{\gamma_{1}\gamma_{2}}{4\pi^{2}}\right]{}_{_{\mathrm{super-fTG}}}\simeq1+\frac{\zeta\left(-1/2\right)a_{\textrm{ho}}}{4\pi\sqrt{N}a_{\perp}}\gamma_{2},\label{eq:BoundarySuper-fTG}
\end{equation}
This estimation - illustrated by a black dashed line in the figure
- agrees qualitatively well with our results and encloses the blue
super-fTG area with small breathing mode frequencies. For a $^{6}$Li
polarized Fermi gas near the $p$-wave Feshbach resonance at $B_{0}=215.0$
G, where $\gamma_{2}\simeq0.14$, we find that $(\gamma_{1})_{\textrm{super-fTG}}\simeq275$
or $(a_{\perp}^{3}/w_{1})_{\textrm{super-fTG}}\simeq-118$. This corresponds
to a detuning from the resonance at about $0.05$ G.

\textit{Summary.} --- We have investigated the thermodynamics and
dynamics of spin-polarized fermions with a resonant $p$-wave interaction
under a one-dimensional harmonic confinement at finite temperature,
by solving the exact Yang-Yang thermodynamic equations and two-fluid
hydrodynamic equation. We have shown that there are distinct features
in the density distribution and collective mode frequency for identifying
an exotic effective-range-induced super fermionic Tonks-Girardeau
state. These features are not sensitive to the presence of a finite
temperature. As a result, our predictions are readily testable with
ultracold $^{6}$Li or $^{40}$K atoms near a $p$-wave Feshbach resonance
at an experimental achievable temperature $T\simeq0.1T_{F}$.
\begin{acknowledgments}
We thank very much Professor Xi-Wen Guan for his explanation on the
Yang-Yang thermodynamics of the 1D $p$-wave Fermi gas with the effective
range of the interaction included. This work was supported by the
ARC Discovery Projects: DP140100637 and FT140100003 (XJL), FT130100815
and DP140103231 (HH).\end{acknowledgments}

\end{document}